*Article*

# Enhancement of nitride-based solar cells using graphene as transparent contact layer

**Miriam Cadenas[1,*], Mireia Martínez[1], Kerly Sánchez[1], Jordi Ibáñez[2], Sergi Hernández[3], Sirona Valdueza-Felip[1], Ana M. Diez-Pascual[4], Fernando B. Naranjo[1]**

[1] Grupo de Ingeniería Fotónica, Universidad de Alcalá, 28871, Alcalá de Henares, Spain
Unidad Asociada Instituto de Óptica-CSIC, Tecnologías Fotónicas y de Sensores; miriam.cadenas@uah.es (M.C.); mireia.martineza@uah.es (M.M.); kerly96sanchez@hotmail.com (K.S.); sirona.valdueza@uah.es (S.V.-F.); fernando.naranjo@uah.es (F.B.N.)
[2] GEO3BCN-CSIC, 08028, Barcelona, Spain; jibanez@geo3bcn.csic.es (J.I.)
[3] MIND, Department of Electronics and Biomedical Engineering, Universitat de Barcelona, 08028, Spain; shernandez@ub.edu (S.H.)
[4] Universidad de Alcalá, Facultad de Ciencias, Departamento de Química Analítica, Química Física e Ingenieria Química, Energía, 28871, Alcalá de Henares, Spain; am.diez@uah.es (A.M.D.-P);
* Correspondence: miriam.cadenas@uah.es (M.C.)

**Abstract:** The effect of using a graphene layer as a semitransparent contact layer is studied in solar cells based on AlInN on Si (100) substrates. The devices consist of $Al_xIn_{1-x}N$ layers deposited on p-type Si (100) substrates incorporating a thin amorphous silicon (a-Si) buffer layer to improve the heterointerface quality. Three aluminum contents are studied, namely: x=0.22, 0.35 and 0.43. Subsequently, a monolayer graphene film was transferred onto the front surface of the devices using a simple and low-temperature transfer process, acting as a semitransparent conductive contact. The photovoltaic characteristics were then evaluated under illumination and dark conditions in devices with and without the graphene layer. The results show that the incorporation of graphene leads to a clear improvement in the short-circuit current density, fill factor, and overall power conversion efficiency for all studied compositions, while the open-circuit voltage remains largely unaffected. These findings demonstrate the potential of graphene as an effective transparent conductive contact for nitride-based solar cells.

Group-III nitride semiconductors have attracted considerable attention for photovoltaic applications due to their wide and tunable direct bandgap, high absorption coefficient, excellent thermal stability, and radiation hardness [1–3]. In particular, the $Al_xIn_{1-x}N$ alloy system allows bandgap engineering from the near-infrared to the deep-ultraviolet by adjusting the aluminum content, making it attractive for high-efficiency single-junction and multijunction solar cells [1,3].

The integration of AlInN with silicon substrates is especially appealing because it combines the advantages of nitride materials with the low cost, large area, and mature processing technology of silicon [4,5]. However, several challenges still limit the performance of AlInN/Si heterojunction solar cells, including the quality of the heterointerface, the presence of defects and dislocations, and the optimization of front and back contacts [6,7].

In a previous work by our group, we shown that the introduction of an amorphous silicon (a-Si) buffer layer of 15 nm between AlInN and the crystalline silicon substrate significantly improve interface quality and device performance, leading to enhanced photovoltaic efficiencies in $Al_xIn_{1-x}N$/Si heterojunction solar cells [8]. The devices investigated in the present work are based on that architecture.

Another key aspect for further performance enhancement is the development of transparent conductive contacts that combine high optical transparency with low sheet resistance and good chemical and mechanical stability. Conventional transparent conductive oxides, such as indium tin oxide, suffer from drawbacks including brittleness, limited transparency in specific spectral ranges, and the requirement of deposition at relatively high temperature [9]. In this context, graphene has emerged as a promising alternative transparent electrode due to its exceptional electrical conductivity, high optical transmittance, mechanical flexibility, and chemical robustness [10–12].

The successful use of graphene as a transparent contact has been demonstrated in various photovoltaic technologies, including silicon, organic, perovskite, and III–V solar cells, where improvements in carrier collection, series resistance, and device stability have been reported [13–16]. Nevertheless, the application of graphene contacts in nitride-based solar cells remains unexplored.

In this work, the effect of introducing a graphene semitransparent contact layer on $Al_xIn_{1-x}$ N/a-Si/Si heterojunction solar cells has been investigated. Three devices with different aluminium contents, 0.22, 0.35 and 0.43, were studied. In all cases, the photovoltaic performance before and after graphene transfer is compared in terms of the current–voltage (J–V) characteristics, including the short-circuit current density (Jsc), open-circuit voltage (Voc), fill factor (FF) and efficiency (Eff), in order to evaluate the role of graphene as a transparent conductive contact in nitride-based photovoltaics.

The $Al_xIn_{1-x}$ N/a-Si/p-Si (100) solar cells investigated in this work are based on the device architecture and fabrication process previously described in reference [8]. Both the a-Si buffer and the AlInN layer were deposited by magnetron reactive sputtering using a combined DC/RF system (AJA International, ATC ORION-3-HV) on commercial p-type Si (100) substrates with a thickness of 375 μm and a resistivity in the range of 1 to 10 Ω·cm. The system is equipped with 2-inch confocal magnetron cathodes using boron-doped Si, In, and Al targets, with the target-to-substrate distance set at 10.5 cm. The pre-deposition base pressure was on the order of $10^{-5}$ Pa.

Before deposition, the Si substrates were chemically cleaned with organic solvents and thermally degassed within the chamber, followed by gentle etching with Ar plasma to remove surface contaminants, as described in reference [8]. The amorphous silicon buffer layer was deposited using a DC-fed Si target in pure Ar plasma at a substrate temperature of 550 °C, with the a-Si buffer thickness set at 15 nm. Subsequently, the AlInN absorber layer was deposited at the same temperature as the substrate using a pure $N_2$ plasma with a fixed RF power of 30 W at the target. The analysis focuses on three devices deposited with aluminium sputtering power of 100, 125, and 150 W, which lead to Al composition in the AlInN absorber layer of 0.22, 0.35 and 0.43, respectively. The nominal thickness of the AlInN layer was approximately 260 nm for all devices.

The characterization of the as prepared materials, including structural, optical, and electrical properties, has already been reported [8]. In particular, the surface morphology analyzed by AFM revealed RMS roughness values in the nanometer range for $Al_xIn_{1-x}N$ films with different aluminum molar fractions (x), with values of approximately 3.6 nm for x = 0.22, 3.5 nm for x = 0.35, and 2.4 nm for x = 0.43, indicating relatively smooth and compact surfaces.

The processing and contact metallization of the solar cells followed the procedure detailed in reference [8]. Briefly, the p-type back contact consisted of an Al layer deposited on the back face of the Si substrate and annealed to ensure ohmic behaviour, while the n-type top contact was formed by metallizing the AlInN surface with Al using a mechanical mask to develop a finger-shape contact. Both back and top contacts have a thickness of 100 nm.

Following electrical characterization of the fabricated devices, a monolayer graphene film was transferred to the front surface of the solar cells to act as a semi-transparent contact layer. This film occupied approximately half the dimensions of the photovoltaic cell's front contact, but spanned all three of its fingers, as shown in Figure 1. Commercially available, easily transferable graphene supported by a PMMA layer was used. Graphene transfer was performed by floating the film in deionized water and then collecting it with the target substrate, followed by drying, mild heat treatment, and PMMA removal in acetone preheated to 50°C. This resulted in a continuous graphene layer covering the active area of the devices, showing no apparent dependence of the surface roughness of the devices, for the analysed range of values.

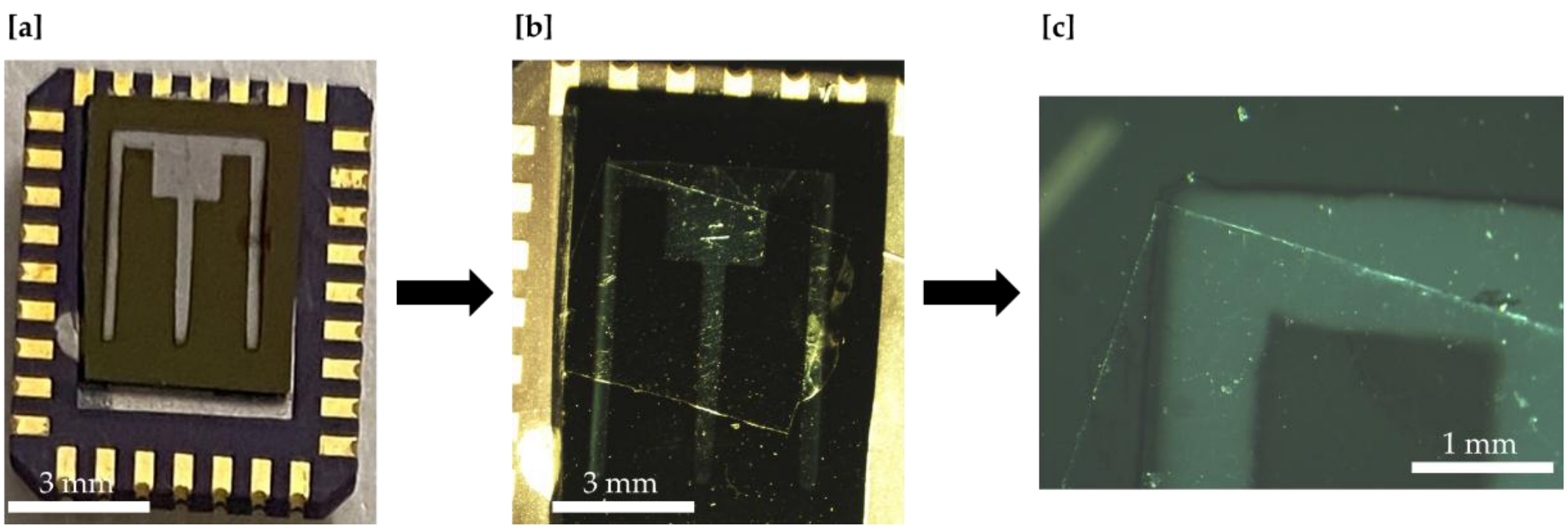

**Figure 1**. Optical images of a representative solar cell with a graphene sheet transferred to its surface at different magnifications: (a) general view of the device fabricated after graphene transfer; (b) magnified view of the same device showing the transferred graphene layer; and (c) image at higher magnification highlighting the graphene sheet on the surface of the device.

The quality of the deposited graphene layer was assessed by Raman measurements, which were excited with the second harmonic of a continuous-wave Nd:YAG laser (λ=532 nm) and acquired with a Horiba Jobin-Yvon LabRam spectrometer coupled to a high-sensitive CCD detector. The light was focused to a ~1 μm diameter spot using a ×100 microscope objective, with an optical power on the sample of ~1 mW.

Figure 2 shows Raman spectra acquired on different regions of the sample, together with a reference Raman spectrum of graphene. The spectra recorded on smooth surface regions of the sample (Point #1) are dominated by broad bands in the 400-900 $cm^{-1}$ region that can be attributed to the AlInN layer. No Raman features attributable to graphene are observed in these smooth regions. By contrast, in addition to the Raman signal from AlInN, the spectra from iridescent regions (Point #2, see the inset of Fig. 2) exhibit two weak features at approximately 1585 and 2680 $cm^{-1}$ that coincide with the G and 2D bands of graphene, respectively. The appearance of these peaks can be attributed to graphene incorporated onto the sample.

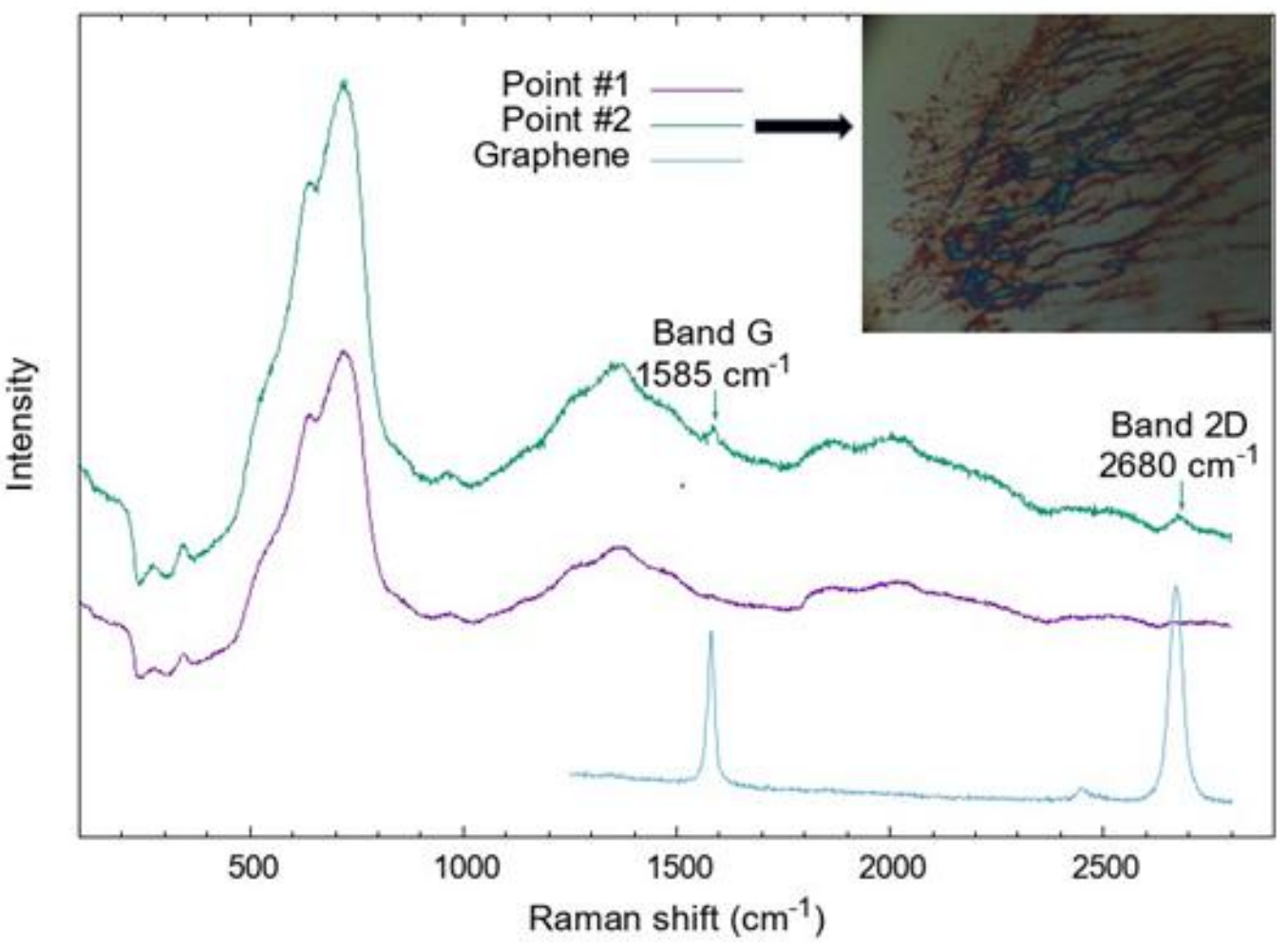

**Figure 2:** Raman spectra from smooth (Point #1) and iridescent (Point #2) regions of the sample. The Raman spectrum from a reference graphene sample is included for comparison.

The electrical and photovoltaic properties of the developed devices were measured following the procedures described in references [17,18]. Current-voltage (J–V) characteristics were obtained both in darkness and under standard AM1.5G illumination using a calibrated solar simulator. This study focused in the important photovoltaic parameters extracted from the J–V curves under illuminated condition: open-circuit voltage (Voc), short-circuit current density (Jsc), fill factor (FF), and power conversion efficiency (η).

Concerning the J-V curves obtained under dark conditions (not shown), the reverse characteristic remained almost unchanged upon the introduction off the graphene layer. On the contrary, an enhancement of the forward characteristics is clearly visible, with a reduction of the series resistant (Rs, given in $\Omega \cdot cm^2$) when introduction of the graphene layer, from 8.6 to 3.4; 6.9 to 4.1; and 7.8 to 5.6; for devices with 0.22, 0.35 and 0.43 Al content, respectively.

Table 1 summarizes the photovoltaic parameters extracted from the J–V curves obtained under illumination conditions of the solar cells before and after the transfer of graphene. The corresponding J–V characteristics for each device with (CG) and without graphene (SG) is shown in Figure 3.

**Table 1**. Photovoltaic parameters extracted from illuminated J–V curves of $Al_xIn_{1-x}$ N/Si solar cells before and after incorporation of a graphene transparent contact. The measurement errors are as follows: $J_{SC}$ (± 0.2), $V_{OC}$ (± 0.02), FF (± 0.25) and Eff. (± 0.03).

| | **Initial** | **With Graphene** | **Initial** | **With Graphene** | **Initial** | **With Graphene** | **Initial** | **With Graphene** |
|---|---|---|---|---|---|---|---|---|
| | **Jsc (mA/cm²)** | | **Voc (V)** | | **FF (%)** | | **Eff. (%)** | |
| **X = 0.22** | -9.3 | -10.0 | 0.44 | 0.45 | 41.43 | 55.18 | 1.72 | 2.53 |
| % Change | - | 7.53 | - | 2.27 | - | 33.19 | - | 47.09 |
| **X = 0.35** | -6.9 | -9.0 | 0.40 | 0.40 | 47.84 | 49.67 | 1.34 | 1.83 |
| % Change | - | 30.43 | - | 0 | - | 3.83 | - | 36.57 |
| **X = 0.43** | -6.6 | -6.8 | 0.35 | 0.35 | 32.16 | 32.49 | 0.76 | 0.79 |
| % Change | - | 3.03 | - | 0 | - | 1.03 | - | 3.95 |

After graphene transfer, the open-circuit voltage remained essentially unchanged, while the short-circuit current density and fill factor showed a noticeable increase. As shown in Figure 3 and summarized in Table 1, this improvement is mainly driven by an increase in Jsc and FF for all compositions, whereas Voc exhibits only minor variations within experimental uncertainty. The J–V curves exhibit an upward shift in the low-voltage region and a steeper slope near the maximum power point, indicating improved carrier collection and reduced series resistance. This effect is particularly evident for the x = 0.22 device deposited at 100 W, which shows the largest relative enhancement in both Jsc and FF after graphene transfer. As a result, the overall power conversion efficiency improved in all cases, being the most pronounced enhancement for the 100 W device. In this case, the post-transfer J–V curve becomes more rectangular, consistent with the simultaneous increase in Jsc and FF.

The improvement can be attributed to the high optical transparency and electrical conductivity of graphene, which allows efficient light transmission while providing an additional conductive path for photogenerated carriers. The reduced slope in the low-voltage region and the improved curve shape near the maximum power point further suggest a decrease in series resistance and enhanced lateral current spreading at the front contact, contributing to the observed performance improvement. These findings are consistent with previous reports of graphene contacts in other photovoltaic systems [13–16], showing its effectiveness in enhancing the performance of solar cell devices.

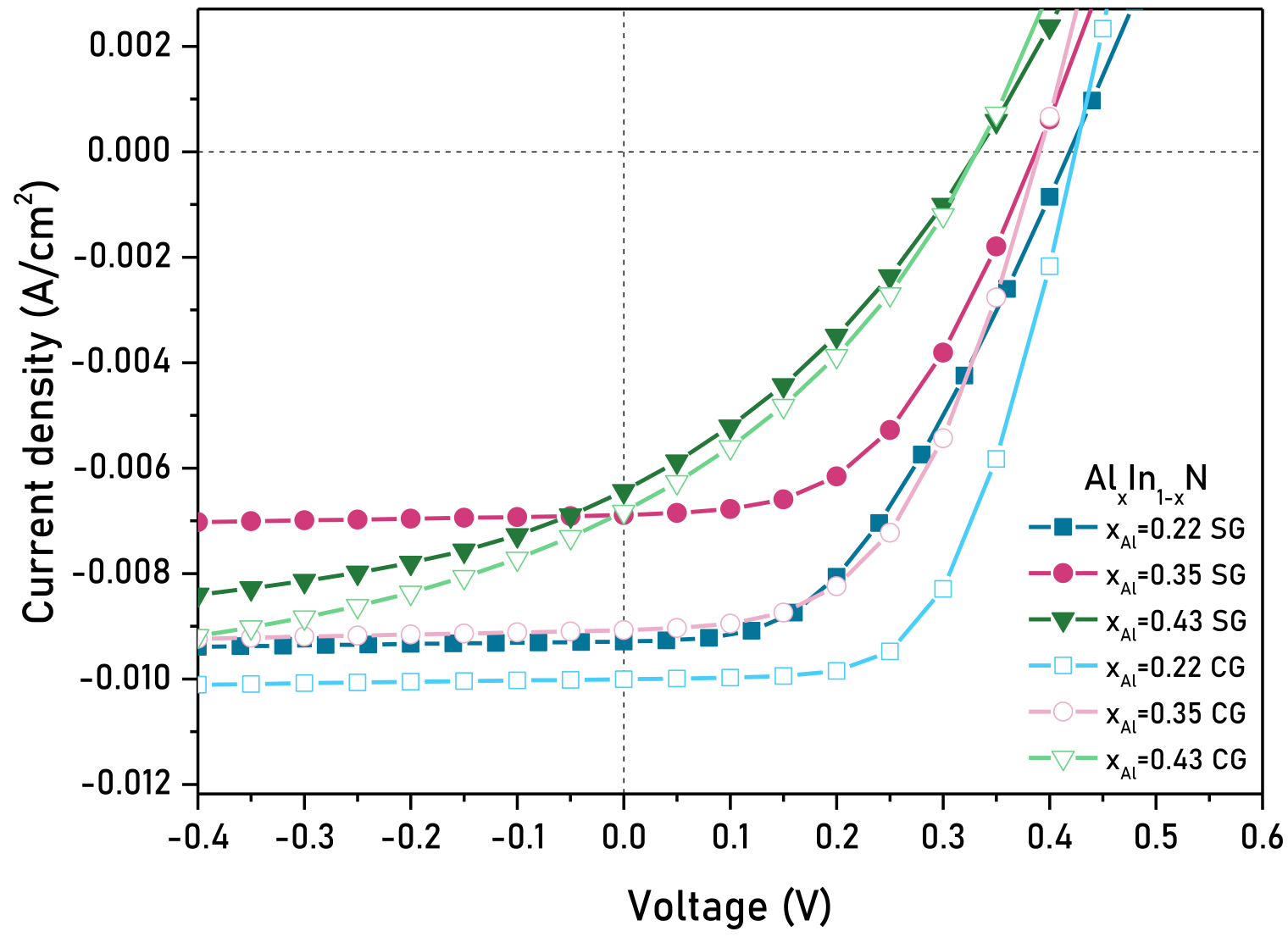

**Figure 3.** J-V curves under AlInN illumination in Si (100) devices with a-Si buffer as a function of Al content without graphene (SG) and with graphene (CG).

**Conclusions**

The incorporation of a graphene semitransparent contact layer in $Al_xIn_{1-x}N$/Si solar cells leads to a reduction of the device series resistance and clear enhancement of key photovoltaic parameters, including short-circuit current density, fill factor, and overall conversion efficiency. The improvement is observed for all the analysed devices, with Al contents of 0.22, 0.35 and 0.45, being the highest performance enhancement for the device with the lowest Al content. In this case, the largest increase of the photovoltaic efficiency of 47,1 % (from 1.72 % to 2.53 % without and with graphene layer, respectively) is achieved, accordingly with the maximum reduction of the series resistance of the device. These results demonstrate that graphene is an effective transparent conductive contact for nitride-based solar cells and can be integrated using a simple, low-temperature transfer process compatible with existing device structures. Further optimization of graphene quality and contact engineering is expected to yield additional performance gains.

**Author Contributions:**

Conceptualization, F.B.N. and A.M. D-P.; methodology, F.B.N. and M. C.; validation, F.B.N. and M.C.; formal analysis, F.B.N. and M.C.; investigation, F.B.N., M.C., M.M. and K.S.; resources, F.B.N. and S.V.-F.; data curation, M.M. and M.C.; Raman measurements and analysis of related results J.I. and S. H.; writing—original draft preparation, F.B.N., M.C., and A.M.D.-P.; writing—review and editing, F.B.N., A.M.D.-P and M.C.; visualization, F.B.N., M.C., and A.M.D.-P.; supervision, F.B.N.

and A.M.D.-P.; project administration, F.B.N. and S.V.-F.; funding acquisition, F.B.N., A.M.D.-P. and S.V.-F.

**Funding:**

This research was funded by the RELY (PDC2023-145888-I00) (MICIU/AEI/10.13039/501100011033 Next Generation EU/PRTR), GRANISENS (PID2024-160429NB-I00), PIUAH24/IA-053 and PTUAH24/016 (UAH) projects.

**Data Availability Statement:**

The original contributions presented in this study are included in the article. Further inquiries can be directed to the corresponding authors.

**Acknowledgments:**

Authors would thank to S. Fernández from CIEMAT for valuable suggestions and discussions.

**Conflicts of Interest:**

The authors declare no conflicts of interest